\documentclass[letterpaper,twocolumn,preprintnumbers,amsmath,amssymb,floatfix,nofootinbib]{revtex4}

\usepackage{graphicx}
\usepackage{bm}
\usepackage{url,multirow,xcolor,longtable}

\begin{document}

\title{ Isoscalar pairing interaction for the quasiparticle random-phase approximation approach to double-$\bm{\beta}$ and $\bm{\beta}$ decays }

\author{J.\ Terasaki}
\affiliation{Institute of Experimental and Applied Physics, Czech Technical University in Prague, Husova 240/5 110$\;$00 Prague 1, Czech Republic}
\author{Y.\ Iwata}
\affiliation{Faculty of Chemistry, Materials and Bioengineering, Kansai University, Yamatemachi 3-3-35, Suita, Osaka, 564-8680, Japan}

\begin{abstract} 
We have proposed in a series of previous papers a method to determine the effective axial-vector current coupling and the strength of the isoscalar proton-neutron pairing interaction for calculating the nuclear matrix elements of the  neutrinoless double-$\beta$ ($0\nu\beta\beta$) decay by the quasiparticle random-phase approximation. The combination of these two parameters have had an uncertainty in the QRPA approach, but now this uncertainty is removed by introducing a mathematical identity derived under the closure approximation to the nuclear matrix element of the $0\nu\beta\beta$ decay. In this paper, we apply our method to the $0\nu\beta\beta$ decays of $^{136}$Xe and $^{130}$Te and show the nuclear matrix elements and reduced half-lives. Our calculation is tested first by a self-check method using the two-neutrino double-$\beta$ decay, and this test ensures the application of our method to $^{136}$Xe. It turns out, however, that our method is not successful in $^{130}$Te. Further tests are made for our calculation, and satisfactory results are obtained for $^{136}$Xe.  \end{abstract}

\pacs{ }
%
\maketitle
\section{\label{sec:introduction}Introduction}
Neutrino physics is in an important era. The neutrino-oscillation experiments have significant progress in terms of precision and accuracy, and nowadays there is a realistic possibility that the hierarchy of the neutrino mass is clarified in the future \cite{NeuPhys2018}. The experiments of the neutrinoless double-$\beta$ ($0\nu\beta\beta$) decay also progress, and the upper limit of the effective neutrino mass (Majorana neutrino mass) is lowered significantly compared to many years ago, e.g., \cite{Gan16}. 

 The nuclear matrix element (NME) of the $0\nu\beta\beta$ decay plays a crucial role for the determination of the effective neutrino mass \cite{Eng17}. This NME and the phase-space factor are necessary for the determination, and these theoretical quantities cannot be confirmed directly experimentally. It seems fair to write that the calculation of the NME is more difficult than the phase-space factor because all of the candidate nuclei of the $0\nu\beta\beta$ decay are heavy, thus, approximation is essential for obtaining the nuclear wave functions. For this reason, an important problem has been how the reliability of the NME calculation can be shown, and if possible, how the calculation can be improved.  

Several theoretical methods or models have been applied for obtaining the nuclear wave functions to calculate the NMEs, and those NMEs are in the  range of a factor of 2$-$3 depending on the method \cite{Eng17}. 
The quasiparticle random-phase approximation (QRPA) approach has a long history of the application to the $0\nu\beta\beta$ decay, e.g., \cite{Eng88,Suh90,Suh91,Suh93,Suh97,Suh98,Fae98,Sim98,Sim99,Bob01,Pac03,Sim04,Alv04,Suh05,Rod07,Kor07,Sim08,Mor09,Sim09,Fan11,Sim13,Mus13,Fan15,Hyv15,Suh17}. 
 Those applications clarified that a crucial point in the calculation procedure is to determine the effective axial-vector current coupling $g_A$ and the strength of the isoscalar, necessarily proton-neutron, pairing interaction, e.g., \cite{Suh17}. Currently phenomenological value is necessary for the former parameter in any approach to the $\beta\beta$ decay. The value of the latter parameter is not well-established in the approaches using phenomenological energy-density functional because the proton-neutron pairing gap is not clear in the experimental data. It was pointed out \cite{Suh17} that the different combinations of the two parameters result in similar NMEs as long as the experimental half-lives in the two-neutrino double-$\beta$ ($2\nu\beta\beta$) decays are reproduced. It would be even better,  if this uncertainty of the combination were removed; the calculation of the NME is a matter of reliability.   

We have proposed \cite{Ter16} a method to solve this problem by introducing a mathematical identity for determining the strength of the isoscalar pairing interaction; no additional symmetry is imposed. The effective $g_A$ is determined so as to reproduce the experimental $2\nu\beta\beta$ half-life; this is as usual in the QRPA approach, e.g., \cite{Sim13}. This method is clear-cut theoretically, however, the obtained value of $g_A$ is $\simeq$ 0.5 much smaller than the usual values $\simeq$ 1.0, e.g., \cite{Sim13}. It would not be surprising, if the appropriate effective $g_A$ depends on the approximation method. Considering the current situation of many $\beta\beta$ studies \cite{Eng17}, it is necessary in our method to examine if our small $g_A$ does not cause any problem. A useful check is the consistency with the $\beta$ decay. As a matter of course, any of the candidate nuclei of the $0\nu\beta\beta$ decay used by the experiments do not have the $\beta$ decay, thus to our knowledge, there has been no theoretical paper on the $\beta\beta$  decay discussing the $\beta$ decay of the parent nucleus simultaneously. In this case, one can study the $\beta$ decay of nuclei close to the candidate nuclei in the nuclear chart assuming that the change in the effective $g_A$ is negligible.

The QRPA is a method to obtain the transition from the ground state to the excited states, e.g., \cite{Rin80,Bla85}. When the pairs of the proton and neutron quasiparticles are used as the building blocks to express the transition, i.e., the proton-neutron QRPA (pnQRPA), the intermediate states of the $\beta\beta$ decay can be represented in two ways; one way is the pnQRPA based on the ground state of the mother nucleus, and another way is to use that of the grand-daughter nucleus. This structure of the dual intermediate-state spaces is a unique feature of the QRPA approach to the $\beta\beta$ decay. Because the QRPA is an approximation, the intermediate state of one space is not identical to a state of another space. We have proposed \cite{Ter18} a simple method of self-check of the validity of the dual intermediate-state spaces by comparing the $2\nu\beta\beta$ NME with the energies of the intermediate states obtained from the ground state of the mother nucleus and that with those energies obtained from the grand-daughter nucleus. The two results are close, if the QRPA is a good approximation. We do this test for $^{136}$Xe $\rightarrow$ $^{136}$Ba and $^{130}$Te $\rightarrow$ $^{130}$Xe in this paper. In addition, we compare the spectra of the intermediate states obtained by the two pnQRPA calculations. This comparison shows how the two intermediate spaces are close.  The purpose of this paper is to show the results of the above tests of the methodology and discuss the validity of our approach. 

Section \ref{sec:calculation_method} is a concise description of the calculation methods of the Hartree-Fock-Bogoliubov (HFB) approximation, QRPA, and NMEs of the $\beta\beta$ decays. The calculation results of the NMEs are shown in Sec.~\ref{sec:results_tests}, and several tests of the calculation are discussed. 
Section \ref{sec:summary} is the summary. 

\section{\label{sec:calculation_method}Calculation method}
\subsection{\label{sec:HFB}HFB calculation}
We proceed according to the original QRPA theory \cite{Rin80,Bla85}. That is, the HFB calculation is performed at the beginning, and the subsequent QRPA calculation is performed using  the HFB ground state and the same Hamiltonian as used in the HFB calculation. 
The HFB code developed according to Refs.~\cite{Ter03,Bla05,Obe07} is used. The quasiparticle wave functions are represented in the two-dimensional cylindrical B-spline mesh with the vanishing boundary condition at the edge of the cylindrical box. The parity and the $z$ component of the angular momentum $j_z$ are good quantum numbers. Only $z>0$ is treated numerically, and the calculation with respect to the angle around the $z$ axis is processed analytically.  These two symmetries are always conserved throughout our calculations below. The maximum $z$ is 20 fm, and the maximum of the radius variable $\rho$ of the direction  perpendicular to the $z$ direction is also 20 fm. The number of the B-spline mesh points is 42 for each coordinate variable with a non-uniform  distribution.

For the Hamiltonian, we use the Skyrme interaction (energy density functional) with the parameter set SkM$^\ast$ \cite{Bar82} and the volume (density independent) contact pairing interactions. The effective single-particle energy range of applying the pairing interaction is from the lowest level up to 60 MeV above the continuum threshold. The nuclear densities and pairing tensors are calculated in this range. The strengths of the pairing interactions for the like-particles are determined so as to reproduce by the HFB calculation the pairing gaps obtained from the experimental masses through the three-point formula \cite{Boh69}. 
The properties of the HFB ground-state solutions of $^{136}$Xe, $^{136}$Ba, $^{130}$Te, and $^{130}$Xe are summarized in Table \ref{tab:HFB_solution}. 
\begin{table}
\caption{\label{tab:HFB_solution} Properties of the HFB solutions of $^{136}$Xe, $^{136}$Ba, $^{130}$Te, and $^{130}$Xe. 
$\Delta_p$ and $\Delta_n$ are the average pairing gaps of the protons and neutrons, respectively, and $\beta$ denotes the quadrupole deformation parameter associated with the quadrupole moment. 
}
\begin{ruledtabular}
\begin{tabular}{cccc}
Nucleus & $\Delta_p$ (MeV) & $\Delta_n$ (MeV) & $\beta$ \\
\hline\\[-10pt]
$^{136}$Xe & $1.341$ & $0$ & $0$ \\
$^{136}$Ba & $1.641$ & $1.158$ & $0$ \\
$^{130}$Te & $1.442$ & $1.359$ & $0$ \\
$^{130}$Xe & $1.692$ & $1.438$ & $0.112$ \\
\end{tabular}
\end{ruledtabular}
\vbox{\vspace{20pt}}
\end{table}

\subsection{\label{sec:QRPA}QRPA calculation}
After the HFB solution is obtained, the canonical single-particle \cite{Rin80} wave functions are obtained by diagonalization of the density. The QRPA equation for the computation is constructed in the so-called matrix formulation \cite{Rin80} with the canonical-quasiparticle basis. 
The dimension of the two-quasiparticle space to define the QRPA equation is truncated by using parameters based on the occupation probabilities of the canonical single-particle states \cite{Ter10}; the physically relevant states are used by this scheme. The dimension is $\simeq$70000 for $K\leq 1$ and less than 40000 for $K\geq 2$ in the like-particle QRPA (lpQRPA) calculations. 
$K$ is the $z$ component of the nuclear angular momentum, and the actual computation is performed only for $K \geq 0$ on the basis of the time-reversal invariance of the ground state. The larger space is used for $K\leq 1$ for the separation of the spurious states from the real states. For the proton-neutron QRPA (pnQRPA), the dimension is always less than 40000. The solutions of $K\leq 8$ were obtained. 

\subsection{\label{sec:calculation_NME}Calculation of $\bm{\beta\beta}$ nuclear matrix element}
 Our unique procedure \cite{Ter16} to obtain the NME of the $\beta\beta$ decay is as follows:
first, the Gamow-Teller (GT) component of the $0\nu\beta\beta$ NME is calculated using the virtual two-particle transfer path [$(Z,N)\rightarrow (Z,N-2)\rightarrow (Z+2,N-2)$, where $Z$ is the proton number, and  $N$ is the neutron number], which is possible under the closure approximation. The lpQRPA is used for obtaining the intermediate states. 
Our numerical solutions of the pnQRPA (see below) do not have a complex-energy solution, thus, the HFB ground state is not in the proton-neutron pair condensation. Therefore, the proton-neutron pairing interaction does not contribute to the  lpQRPA calculation. The spurious states inherent to the lpQRPA are not included in the calculation of the NME. 

The $0\nu\beta\beta$ NME is also obtained by the original $\beta\beta$ path [$(Z,N)\rightarrow (Z+1,N-1)\rightarrow (Z+2,N-2)$] using the pnQRPA. The average of the proton-proton and neutron-neutron pairing interactions is used for the strength of the isovector proton-neutron pairing interaction assuming the isospin invariance of the system. The strength of the isoscalar pairing interaction is determined so as to reproduce the GT component of the $0\nu\beta\beta$ NME obtained by the virtual-path calculation. The equivalence of the two decay paths is a constraint on the effective interactions  for the QRPA. The assumption for this step is that the effective interactions other than the isoscalar pairing interaction are established. The strengths of the pairing interactions are summarized in Table \ref{tab:strength_pairing_interaction}. The Fermi component of the NME is  controlled mainly by the isovector proton-neutron pairing interaction. The tensor component \cite{Doi85} of the NME is neglected throughout our paper because it is known that this component is relatively small. 

Next, the $2\nu\beta\beta$ NME is calculated by the pnQRPA with those strengths of the pairing interactions, and the effective $g_A$ is determined so as to reproduce the experimental half-life in the $2\nu\beta\beta$ decay. 
Finally, the $0\nu\beta\beta$ NME is calculated with this $g_A$ and the pnQRPA solutions already obtained. This $g_A$ is a choice, and the result with the bare value $g_A=1.27$ is also shown later. 
We checked the convergence of the result with respect to the dimension of the single-particle spaces (see below).  Therefore, our calculation is less uncertain than calculations without this check. 

\begin{table}
\caption{\label{tab:strength_pairing_interaction} Strengths of proton-proton ($G_{pp}$), neutron-neutron ($G_{nn}$), isovector proton-neutron ($G_{pn}^\mathrm{IV}$), and isoscalar proton-neutron ($G_{pn}^\mathrm{IS}$) pairing interactions. The unit is MeV$\,$fm$^3$ for all strengths. }
\begin{ruledtabular}
\begin{tabular}{ccccc}
Nucleus & $G_{pp}$ & $G_{nn}$ & $G_{pn}^\mathrm{IV}$ & $G_{pn}^\mathrm{IS}$ \\
\hline\\[-10pt]
$^{136}$Xe & $-194.3$ & $-179.9$ & $-187.1\,\,\,$ & $-55.0$ \\
$^{136}$Ba & $-200.5$ & $-189.3$ & $-194.9\,\,\,$ & $-55.0$ \\
$^{130}$Te & $-219.8$ & $-179.9$ & $-199.85$ & $-50.0$ \\
$^{130}$Xe & $-219.8$ & $-181.8$ & $-200.8\,\,\,$ & $-50.0$ \\
\end{tabular}
\end{ruledtabular}
\vbox{\vspace{20pt}}
\end{table}

\vfill

\section{\label{sec:results_tests} Calculation results and tests }
\subsection{\label{sec:2v}Self-check using $\bm{2\nu\beta\beta}$ decay}
Two NMEs of the $2\nu\beta\beta$ decay, $M^{(2\nu)}(I)$ and $M^{(2\nu)}(F)$, are calculated by\footnote{The notation of Ref.~\cite{Doi85} is used. It is also used for the $0\nu\beta\beta$ NME.}  
\begin{eqnarray}
M^{(2\nu)}(I) &=& \frac{ M^{(2\nu)}_{GT}(I) }{ \mu_0 } - \frac{ g_V^2 }{ g_A^2 } \frac{ M^{(2\nu)}_F (I) }{ \mu_{0F} }, \label{eq:M2v(I)}
\end{eqnarray}
\begin{eqnarray}
M^{(2\nu)}(F) &=& \frac{ M^{(2\nu)}_{GT}(F) }{ \mu_0 } - \frac{ g_V^2 }{ g_A^2 } \frac{ M^{(2\nu)}_F (F) }{ \mu_{0F} }, \label{eq:M2v(F)}
\end{eqnarray}
\begin{eqnarray}
\frac{ M^{(2\nu)}_{GT}(I) }{ \mu_0 } &=& \sum_{K=0,\pm 1} \sum_{a^K_I, a^K_F}
\frac{1}{ \mu_a (I) } \langle F | \tau^- (-)^K \sigma_{-K} | a^K_F\rangle \nonumber \\
&&\times \langle a^K_F | a^K_I \rangle \langle a^K_I | \tau^- \sigma_K | I \rangle , \label{eq:M2vGT(I)}
\end{eqnarray}
\begin{eqnarray}
\frac{ M^{(2\nu)}_{GT}(F) }{ \mu_0 } &=& \sum_{K=0,\pm 1} \sum_{a^K_I, a^K_F}
\frac{1}{ \mu_a (F) } \langle F | \tau^- (-)^K \sigma_{-K} | a^K_F\rangle \nonumber \\
&&\times \langle a^K_F | a^K_I \rangle \langle a^K_I | \tau^- \sigma_K | I \rangle , \label{eq:M2vGT(F)}
\end{eqnarray}
\begin{eqnarray}
\frac{ M^{(2\nu)}_F (I) }{ \mu_{0F} } &=& \sum_{a_I, a_F} \frac{1}{ \mu_a (I) } \langle F | \tau^- | a_F \rangle 
\langle a_F | a_I \rangle \langle a_I | \tau^- | I \rangle, \nonumber \\
\label{eq:M2vF(I)}
\end{eqnarray}
\begin{eqnarray}
\frac{ M^{(2\nu)}_F (F) }{ \mu_{0F} } &=& \sum_{a_I, a_F} \frac{1}{ \mu_a (F) } \langle F | \tau^- | a_F \rangle 
\langle a_F | a_I \rangle \langle a_I | \tau^- | I \rangle, \nonumber \\
\label{eq:M2vF(F)}
\end{eqnarray}
\begin{eqnarray}
\mu_a (I) &=& \frac{1}{ m_e c^2 } \left( E_{aK}^I - \bar{M} \right) , \label{eq:mean_M(I)}
\end{eqnarray}
\begin{eqnarray}
\mu_a (F) &=& \frac{1}{ m_e c^2 } \left( E_{aK}^F - \bar{M} \right) . \label{eq:mean_M(F)}
\end{eqnarray}
$|I\rangle$ and $|F\rangle$ are the ground states of the initial and final nuclei, respectively. 
The intermediate states are denoted by $|a^K_I\rangle$ (obtained from $|I\rangle$) and $|a^K_F\rangle$ (obtained from $|F\rangle$). Those used for the Fermi NME [Eqs.~(\ref{eq:M2vF(I)}) and (\ref{eq:M2vF(F)})] are the  $K=0$ states, and the $K$ is omitted. The  operator changing a neutron to proton is denoted by $\tau^-$, and $\sigma_K$ is the spherical component of the spin-Pauli matrix operator; $\tau^-\sigma_K$ in the equations of the GT component [Eqs.~(\ref{eq:M2vGT(I)}) and (\ref{eq:M2vGT(F)})] is a one-body operator, and $\tau^-$ in Eqs.~(\ref{eq:M2vF(I)})  and (\ref{eq:M2vF(F)}) is another one-body operator. $E^I_{aK}$ and $E^F_{aK}$ are the energies of the intermediate states obtained from $|I\rangle$ and $|F\rangle$, respectively. $\bar{M}$ is the mean value of the masses of the initial and final nuclei, and $m_ec^2$ is the electron mass. The parameter $g_V$ is the vector-current coupling, which is always equal to 1 in our calculations. 
The overlap $\langle a_F^K | a_I^K \rangle$ is calculated according to the method of Refs.~\cite{Ter12,Ter13,Ter15}; the many-body correlations are included by an expansion-truncation. 
The only difference between $M^{(2\nu)}(I)$ and $M^{(2\nu)}(F)$ is which energy ($E^I_{aK}$ or  $E^F_{aK}$) is used in the energy denominator. Usually, the mean value of $E^I_{aK}$ and $E^F_{aK}$ is used. If the higher-order many-body effects beyond the QRPA is small, we will have 
\begin{eqnarray}
M^{(2\nu)}(I) \approx M^{(2\nu)}(F) . \label{eq:M2v(I)approxM2V(F)}
\end{eqnarray}
This is the check point of the validity of the QRPA approach to the $2\nu\beta\beta$ decay. 

The half-life in this decay is calculated by  
\begin{eqnarray}
T_{1/2}^{(2\nu)} &=& \frac{1}{ G_{2\nu}^{(0)} g_A^4 | M^{(2\nu)}|^2}, 
\end{eqnarray}
where $G_{2\nu}^{(0)}$ is the phase-space factor, and we refer to the values in Ref.~\cite{Kot12}. 
Table \ref{tab:NME_2vbb} shows the results of calculation of these decays of $^{136}$Xe and $^{130}$Te. 
For $^{136}$Xe $\rightarrow$ $^{136}$Ba, the same value of $g_A$ can be used for the two calculations to reproduce the experimental half-life, thus, the QRPA is good for this decay instance. On the other hand, the  $g_A$ for fitting the $T^{(2\nu)}_{1/2}$ depends on the choice of the set of the intermediate-state energy for $^{130}$Te $\rightarrow$ $^{130}$Xe. This decay instance turns out to be an example for which the QRPA is not good. 
The absolute values of $M^{(2\nu)}_F/\mu_{0F}$ are much smaller than those of $M^{(2\nu)}_{GT}/\mu_0$ reflecting on the approximate isospin invariance of the isovector pairing interaction. 
\begin{table*}[]
\caption{\label{tab:NME_2vbb} Choice of set of intermediate-state energies, NMEs of the $2\nu\beta\beta$ decay, its components, $g_A$, the calculated half-life in $2\nu\beta\beta$ decay $T_{1/2}^{(2\nu)}$, and the corresponding experimental half-life $T_{1/2}^{(2\nu)}$(exp) \cite{Bar13}. $M^{(2\nu)}$ implies $M^{(2\nu)}(I)$ or $M^{(2\nu)}(F)$, and the same abbreviation is applied to other NMEs and $T_{1/2}^{(2\nu)}$.}
\begin{ruledtabular}
\begin{tabular}{cccccccc}
\multirow{2}{*}{Initial nucleus} & Intermediate- & \multirow{2}{*}{$M^{(2\nu)}$} & \multirow{2}{*}{$\frac{M^{(2\nu)}_{GT}}{\mu_0}$} & \multirow{2}{*}{$\frac{M^{(2\nu)}_{F}}{\mu_{0F}}$} &  \multirow{2}{*}{$g_A$} & $T_{1/2}^{(2\nu)}$ & $T_{1/2}^{(2\nu)}$(exp)\\
& \raisebox{3pt}{state energy} &  &  &  &  & ($10^{21}$yr) & ($10^{21}$yr)\\
\hline
\multirow{2}{*}{$^{136}$Xe} & $E^I_{aK}$ & $0.0748$ & $0.0673$ & $-0.00181$ & $0.49$ & $2.164$ & \multirow{2}{*}{$2.20\pm0.06$} \\
& $E^F_{aK}$ & $0.0744$ & $0.0666$ & $-0.00188$ & $0.49$ & $2.188$ & \\
\multirow{2}{*}{$^{130}$Te} & $E^I_{aK}$ & $0.133\,\,\,$ & $0.132\,\,\,$ & $-0.00029$ & $0.48$ & $0.697$ & \multirow{2}{*}{$0.69\pm 0.13$}\\
& $E^F_{aK}$ & $0.0864$ & $0.0856$ & $-0.00028$ & $0.60$ & $0.677$ & 
\end{tabular}
\end{ruledtabular}
\end{table*}

\subsection{\label{sec:spectrum}Spectrum of intermediate nucleus}
\begin{figure*}[]
\centering
\begin{minipage}[]{.48\textwidth}
\includegraphics[width=1.0\columnwidth]{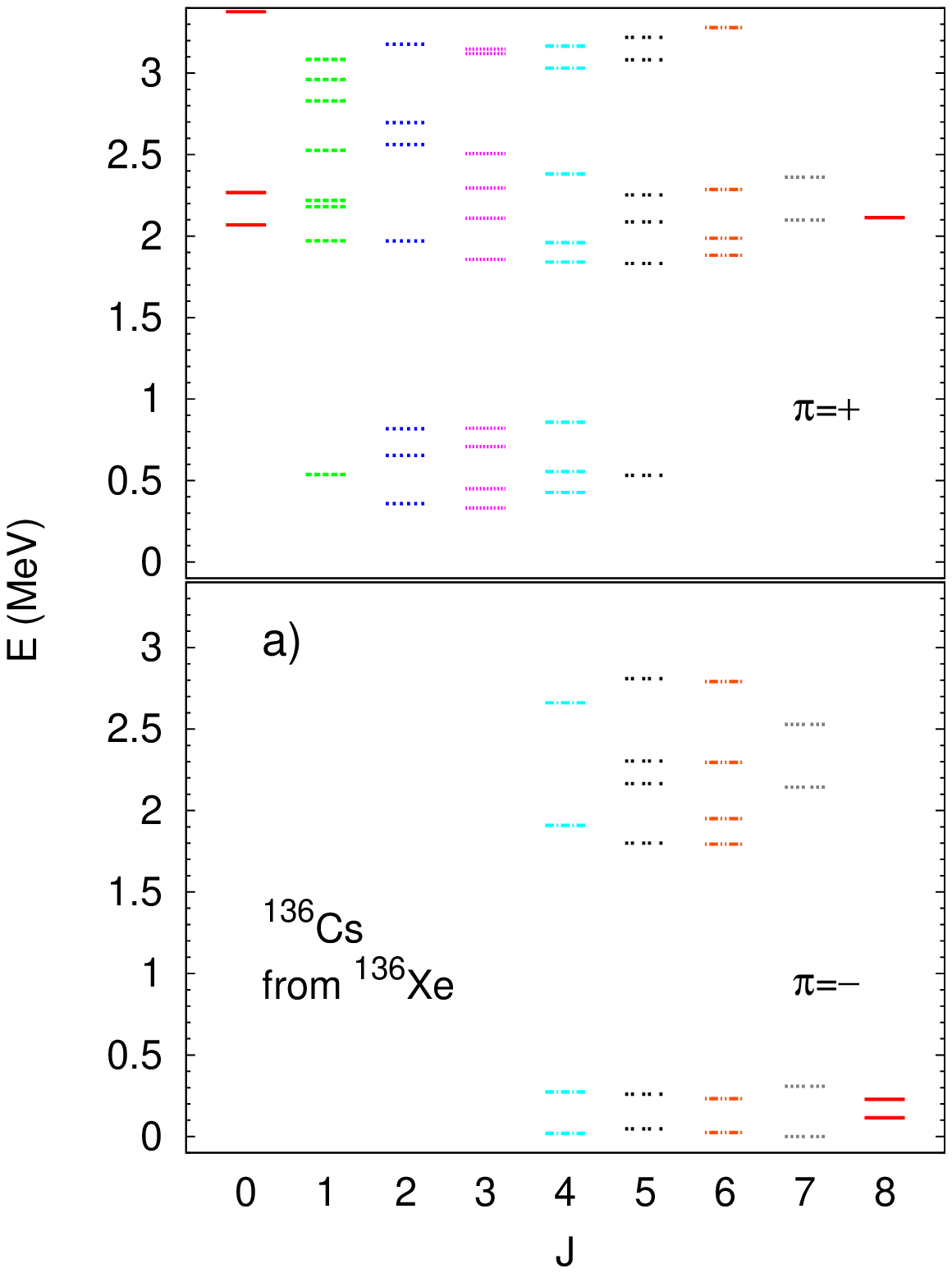}
\end{minipage}
\begin{minipage}[]{.48\textwidth}
\includegraphics[width=1.0\columnwidth]{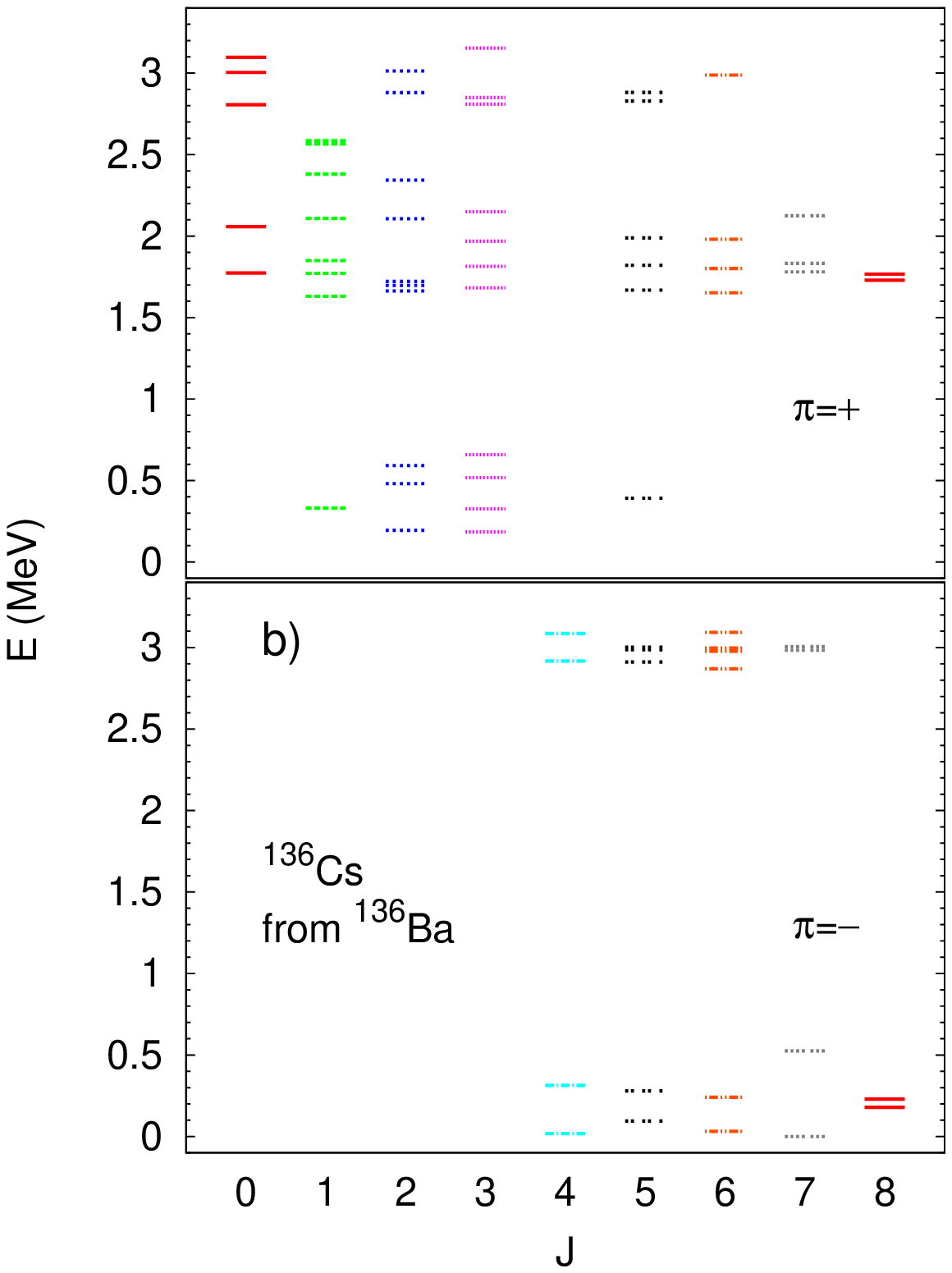}
\end{minipage}
\caption{ \label{fig:spectrum_136Cs_cal} Spectra of $^{136}$Cs in a low-energy region, a) obtained based on the ground state of $^{136}$Xe (including upper panel) and b) obtained from $^{136}$Ba (also upper panel). The upper (lower) panels show the levels of the positive (negative) parity. $J$ is the angular momentum.}
\end{figure*}

 The calculated spectra of the intermediate nucleus $^{136}$Cs are presented in Fig.~\ref{fig:spectrum_136Cs_cal} to further show the validity of the QRPA approach to the decay of $^{136}$Xe. The left spectrum is obtained by the pnQRPA calculation based on $^{136}$Xe, and the right one is obtained based on $^{136}$Ba. We used the transition strength for identifying the intermediate nucleus; see Appendix \ref{sec:appendix1}. Overall, the two results are close to each other. States of angular momentum $J \leq 3$ and $\pi=-$ were not found in the shown energy region. There is a major  gap of levels around 1 MeV for any $J^\pi$. The correspondence of the levels can also be found in many parts between the two calculations. There are two major differences. There is no $J^\pi=4^+$ level in the right figure. This is because there is no transition probability from the ground state of $^{136}$Ba.  The levels around 3 MeV in the right lower figure are more tightly gathered than those in the left lower figure. The satisfactory result of the check using the $2\nu\beta\beta$ decay is endorsed by these spectra. 

\begin{figure}[]
\includegraphics[width=1.0\columnwidth]{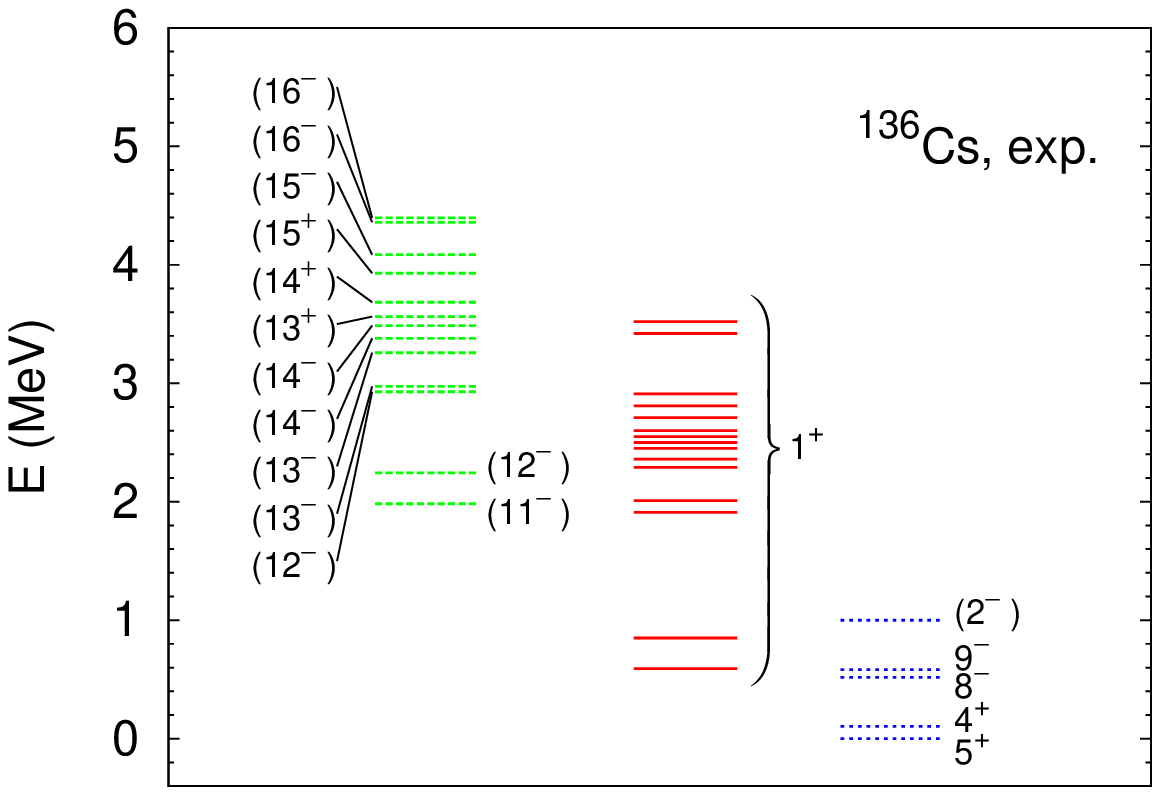}
\caption{ \label{fig:spectrum_136Cs_exp} Observed energy levels of $^{136}$Cs with $J^\pi$ in the discrete-energy region \cite{nndc}. The $J^\pi$'s in parentheses are tentatively assigned.}
\end{figure}

The latest status of the observed states \cite{nndc} is illustrated in Fig.~\ref{fig:spectrum_136Cs_exp}. A gap is seen between 1.0 and 1.9 MeV, and no low-$J$  negative-parity level is seen except for one to which 2$^-$ is assigned tentatively. The $1^+$ levels are observed through the $\beta$ decays, and the high-$J$ levels higher than 2.0 MeV are observed through fissions. Apparently, what states are observed depends on the experimental method. Thus, it is speculated that not all states have  been observed in the discrete-energy region. The features of the calculated spectrum seem to be seen in the experimental data, although this agreement is not conclusive. The ground state is a $7^-$ state in our calculation, but it is actually a $5^+$ state  in the experimental data. 
It is noted that the Skyrme parameter set SkM$^\ast$ is constructed so as to reproduce the properties of the even-even nuclear ground states on average in a broad region of the nuclear chart with emphasis on the doubly-magic nuclei, and fission barrier is also taken into account. Properties of odd-odd nuclei are not taken into account at all. 

\begin{figure*}[]
\centering
\begin{minipage}[]{.48\textwidth}
\includegraphics[width=1.0\columnwidth]{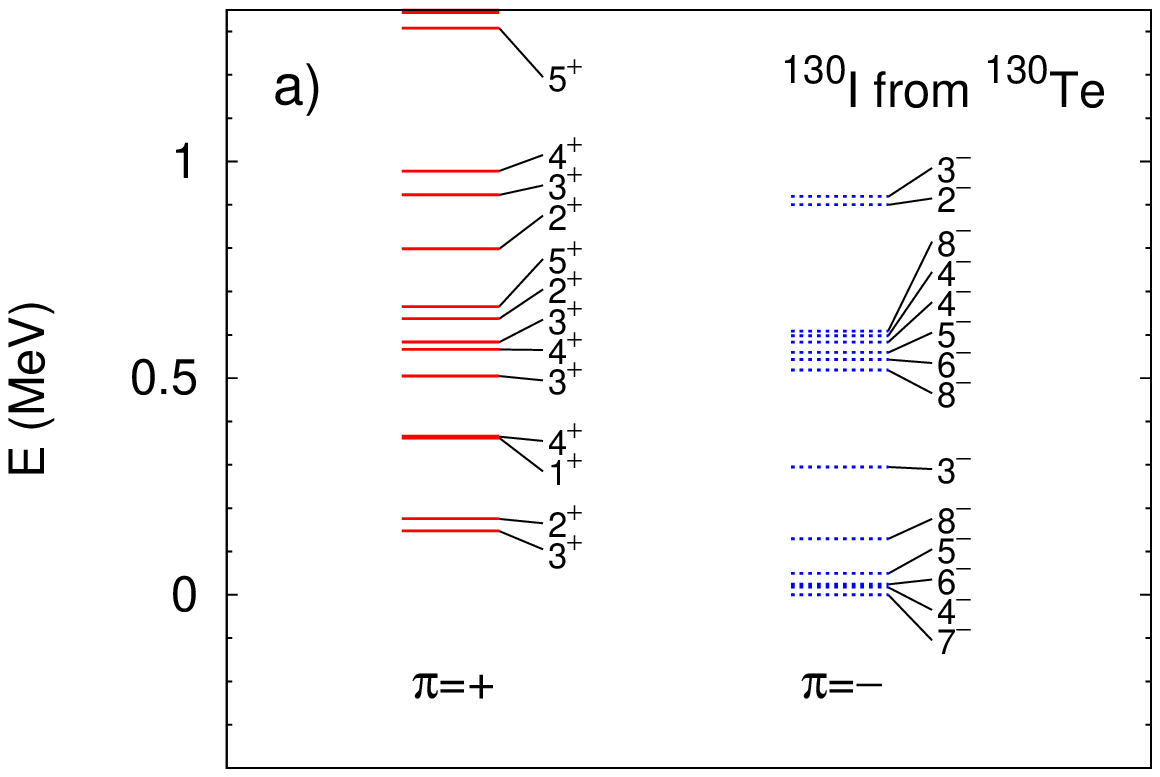}
\end{minipage}
\begin{minipage}[]{.48\textwidth}
\includegraphics[width=1.0\columnwidth]{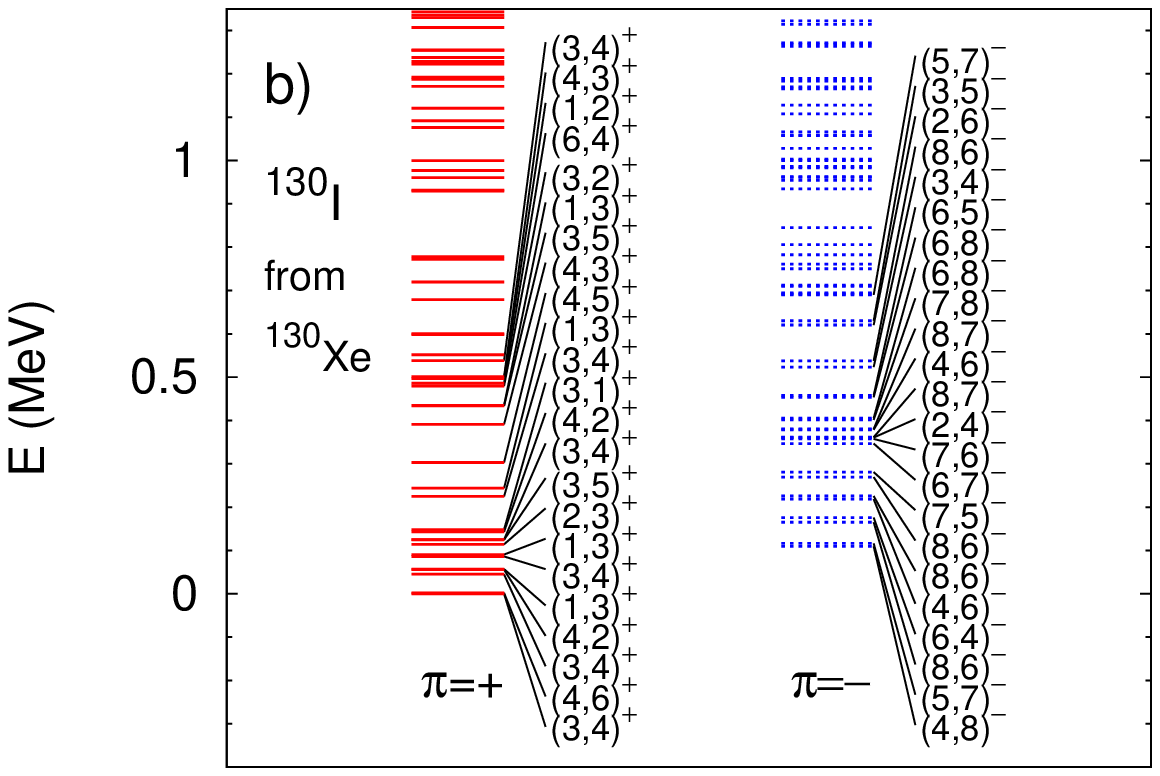}
\end{minipage}
\caption{ \label{fig:spectrum_130I_cal} Spectrum of $^{130}$I with $J^\pi$ in a low-energy region, a) that obtained based on the ground state of $^{130}$Te and b) that obtained from $^{130}$Xe.  Two $J$'s are assigned to each level in the low-energy region of panel b) because the HFB ground state of $^{130}$Xe is deformed; see Appendix \ref{sec:appendix2}.}
\end{figure*}

 We also show the calculated spectrum of $^{130}$I in Fig.~\ref{fig:spectrum_130I_cal}. Panels a and b show the levels obtained from $^{130}$Te and those obtained from $^{130}$Xe, respectively. The level density is clearly different between the two results. The origin of this difference is that the HFB ground state of $^{130}$Te is spherical, but that of $^{130}$Xe is deformed (see Table \ref{tab:HFB_solution}). 
The QRPA has a mathematical property that the broken symmetry in the HFB ground state is restored,  e.g., \cite{Rin80}, however, this property is limited in the QRPA order. In our experience with the like-particle excitations of deformed nuclei \cite{Ter10, Ter11}, $J$ is clear in the states with large  transition strengths, but mixture is seen in the states with small transition strengths.
The deformation implies disappearance of the degeneracy of spherical states. 
Thus, Fig.~\ref{fig:spectrum_130I_cal} indicates that those states obtained from $^{130}$Xe are deformed. Therefore, the two most major values of $J$ are assigned to each level in panel b; for the method to assign $J$ to the deformed states, see Appendix \ref{sec:appendix2}. The  $\beta\beta$ decay of $^{130}$Te is disadvantageous to the QRPA approach due to the difference between the two spectra. 

\begin{figure}[]
\includegraphics[width=1.0\columnwidth]{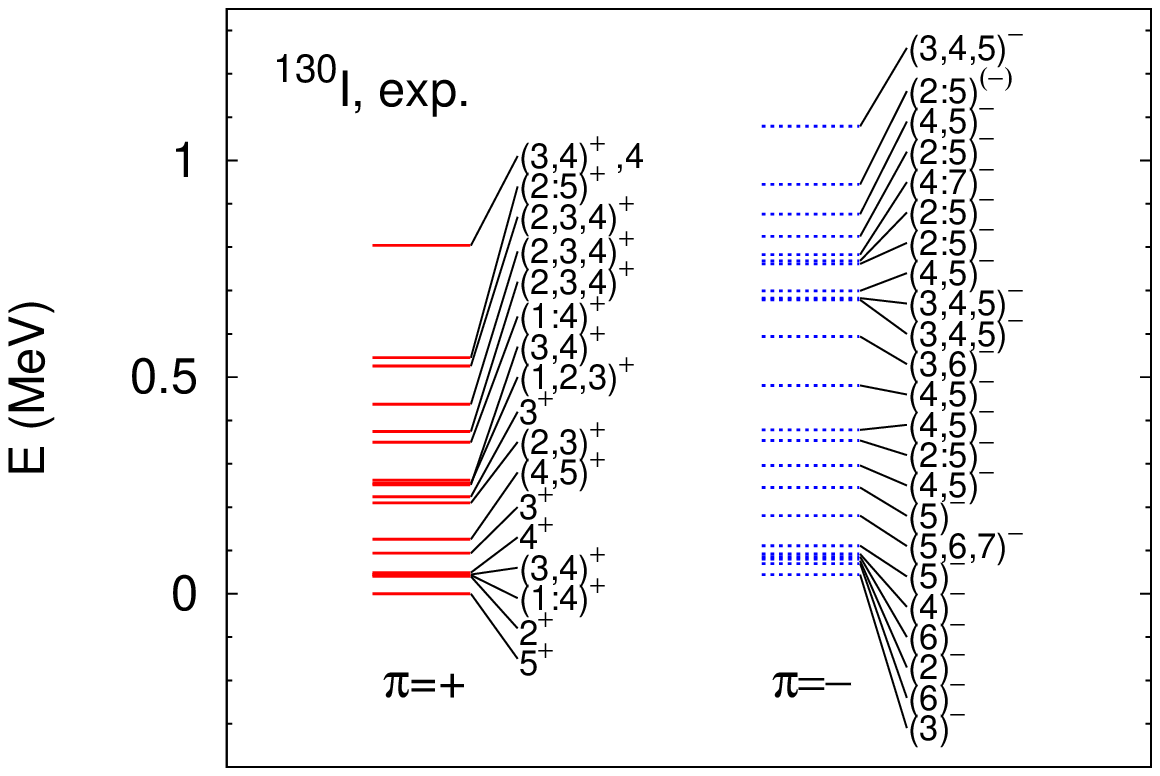}
\caption{ \label{fig:spectrum_130I_exp} Observed energy levels of $^{130}$I with $J^\pi$ in the discrete-energy region \cite{nndc}. The $J^\pi$'s in parentheses are tentatively assigned.}
\end{figure}

Figure \ref{fig:spectrum_130I_exp} illustrates most of the experimental levels compiled in Ref.~\cite{nndc}. The experimental spectrum is closer to that obtained from $^{130}$Xe. Thus, we speculate that the experimental data indicate the deformation effect, assuming that the many-particle-many-hole correlations do not have significant effects to increase the level density in the low-energy region.  

\subsection{\label{sec:0v}$\bm{0\nu\beta\beta}$ decay}
The $0\nu\beta\beta$-decay NME $M^{(0\nu)}$ is calculated by 
\begin{eqnarray}
M^{(0\nu)} &=& M^{(0\nu)}_{GT} - \frac{ g_V^2 }{ g_A^2 } M^{(0\nu)}_F , \label{eq:M0v}
\end{eqnarray}
\begin{eqnarray}
M^{(0\nu)}_{GT} &=& \sum_K \sum_{a^K_I a^K_F} \sum_{pnp^\prime n^\prime}
V^{GT(0\nu)}_{pp^\prime,nn^\prime}(\bar{E}_a) \langle F | c_p^\dagger c_n | a^K_F \rangle \nonumber\\
&&\times\langle a^K_F | a^K_I \rangle\langle a^K_I | c_{p^\prime}^\dagger c_{n^\prime} | I \rangle , \label{eq:M0vGT}
\end{eqnarray}
\begin{eqnarray}
M^{(0\nu)}_{F} &=& \sum_K \sum_{a^K_I a^K_F} \sum_{pnp^\prime n^\prime}
V^{F(0\nu)}_{pp^\prime,nn^\prime}(\bar{E}_a) \langle F | c_p^\dagger c_n | a^K_F \rangle \nonumber\\
&&\times\langle a^K_F | a^K_I \rangle\langle a^K_I | c_{p^\prime}^\dagger c_{n^\prime} | I \rangle , \label{eq:M0vF}
\end{eqnarray}
where $p$ and $p^\prime$ denote the protons, and $n$ and $n^\prime$ denote the neutrons. 
Operators $c^\dagger_i$ and $c_i$ denote the creation and annihilation operators of particle $i$, respectively. 
The two-body transition matrix elements are defined by
\begin{eqnarray}
\lefteqn{ V^{GT(0\nu)}_{pp^\prime,nn^\prime}(\bar{E}_a) } \nonumber\\
 &=& \langle pp^\prime | h_+(r_{12},\bar{E}_a) \bm{\sigma}(1)\cdot \bm{\sigma}(2) \tau^-(1)\tau^-(2) | nn^\prime\rangle , \label{eq:V0vGT} 
\end{eqnarray}
\begin{eqnarray}
V^{F(0\nu)}_{pp^\prime,nn^\prime}(\bar{E}_a) = \langle pp^\prime | h_+(r_{12},\bar{E}_a) \tau^-(1)\tau^-(2) | nn^\prime\rangle . \label{eq:V0vF}
\end{eqnarray}
The argument of $\bm{\sigma}$ and $\tau^-$ distinguishes the two particles that the operators act on. 
We use the following equation for the neutrino potential:
\begin{eqnarray}
h_+(r_{12}, \bar{E}_a) &\simeq& \frac{R}{ r_{12} } \frac{2}{\pi}\left\{ \sin\left( \frac{c}{\hbar}\bar{\mu}_a m_e r_{12}\right) 
\mathrm{ci}\left( \frac{c}{\hbar} \bar{\mu}_a m_e r_{12} \right) \right. \nonumber \\
&&\!\!\!\!\left. -\cos\left( \frac{c}{\hbar} \bar{\mu}_a m_e r_{12} \right) \mathrm{si}\left( \frac{c}{\hbar} \bar{\mu}_a m_e r_{12} \right) \right\} , 
\label{eq:v-potential}
\end{eqnarray}
\begin{eqnarray}
\bar{\mu}_a &=& \frac{1}{ m_e c^2 }(\bar{E}_a - \bar{M} ) .
\end{eqnarray}
This neutrino potential is derived by neglecting the effective neutrino mass compared to the major momentum transfer by the propagating neutrino \cite{Doi85}. 
$R$ is the root-mean-square radius of the nucleus, $r_{12}$ is the distance between two particles, and $\bar{E}_a$ is the average energy introduced in the closure approximation. $R=1.2A^{1/3}$ fm, with the mass number $A$,  and $\bar{\mu}_a=18.51$ are used in our calculations. 
In Eq.~(\ref{eq:v-potential}), functions
\begin{eqnarray}
\mathrm{si}(x) &=& -\int_x^\infty \frac{\sin(t)}{t}dt , 
\end{eqnarray}
\begin{eqnarray}
\mathrm{ci}(x) &=& -\int_x^\infty \frac{\cos(t)}{t}dt ,
\end{eqnarray}
are used. 
Our simple neutrino potential does not include the effects of the dipole form factors \cite{Sim99}, which reduces the $0\nu\beta\beta$ NME by $20-30$ \%, the short-range correlations \cite{Sim09} also reducing the NME, and other subleading terms. 
Our $0\nu\beta\beta$ NME could be overestimated by up to, or at least, 50 \%. 

We calculate reduced half-life in the $0\nu\beta\beta$ decay
\begin{eqnarray}
R^{(0\nu)}_{1/2} = \frac{ m_e^2 c^4 }{ G_{0\nu}g_A^4 |M^{(0\nu)}|^2 } , \label{eq:reduced_half-life}
\end{eqnarray}
where $G_{0\nu}$ is the phase-space factor of the $0\nu\beta\beta$ decay. 
$R^{(0\nu)}_{1/2}$ is the quantity necessary for determining the effective neutrino mass $\langle m_\nu \rangle$ as seen from
\begin{eqnarray}
\langle m_\nu \rangle^2 = \frac{ R^{(0\nu)}_{1/2} }{ T^{(0\nu)}_{1/2} }, 
\end{eqnarray}
with the half-life in the $0\nu\beta\beta$ decay $T^{(0\nu)}_{1/2}$. 
The correct approximations should give the same $R^{(0\nu)}_{1/2}$, if different effective $g_A$ values are  used, because $\langle m_\nu \rangle$ and $T^{(0\nu)}_{1/2}$ are unique.\footnote{ $M^{(0\nu)\prime}=[g_A^2/(g_A^\mathrm{bare})^2]M^{(0\nu)}$ is also used sometimes in the literatures, e.g., \cite{Sim08},  with the bare value of $g_A^\mathrm{bare} \approx 1.27$. } 

\begin{figure}[]
\includegraphics[width=1.0\columnwidth]{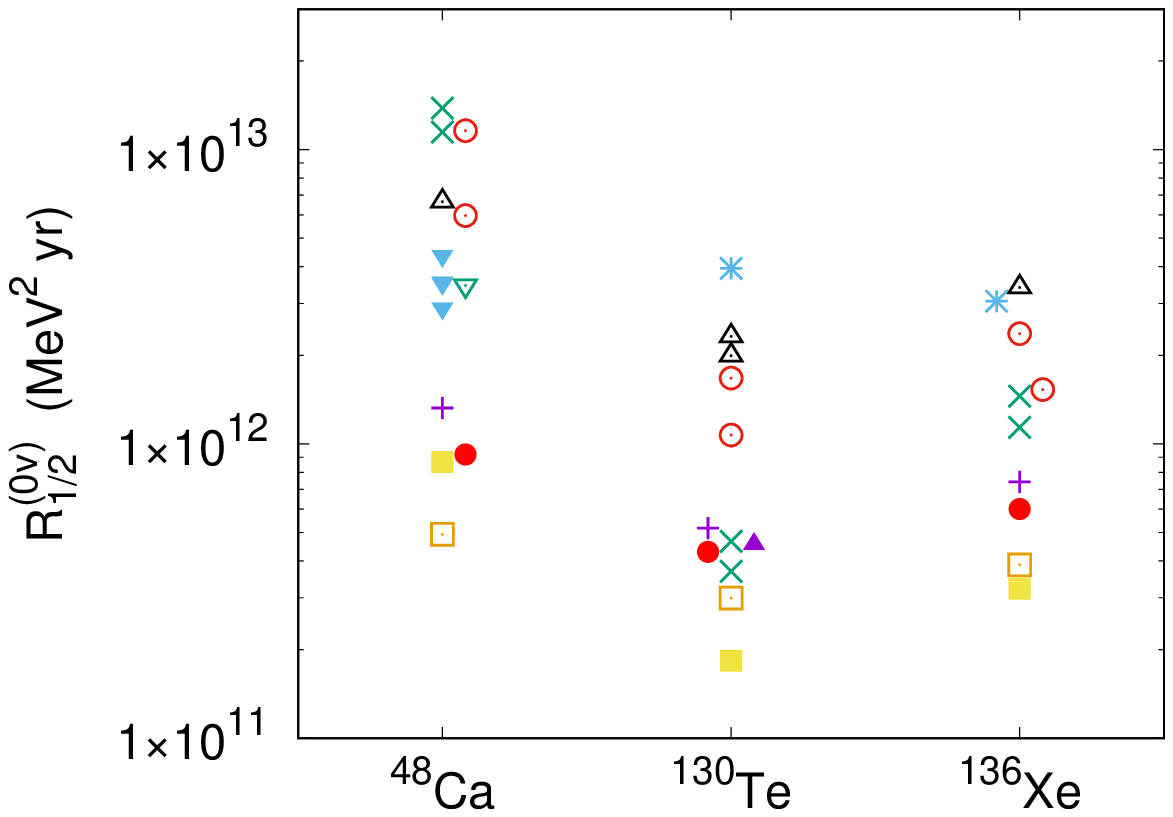}
\includegraphics[width=1.0\columnwidth]{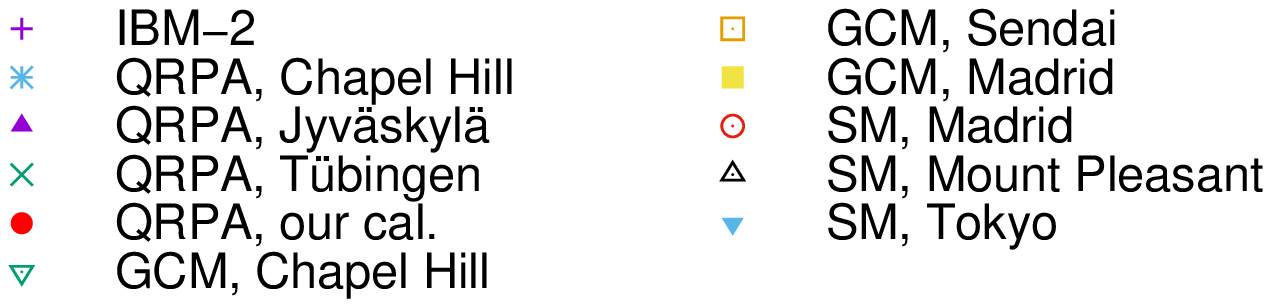}
\caption{ \label{fig:reduced_half-life} $R^{(0\nu)}_{1/2}$ calculated by different groups for $^{48}$Ca, $^{130}$Te, and $^{136}$Xe with $g_A = 1.25$$-$$1.27$. 
SM, GCM, and IBM stand for shell model, generator-coordinate method, and interacting-boson model, respectively. The group of Ref.~\cite{Vaq13} calls their method energy-density functional. 
The references and values of the used $g_A$ are as follows: $^{48}$Ca, \cite{Sim13}, 1.27 (QRPA, T\"{u}bingen); \cite{Hor16}, 1.254 (SM, Mount Pleasant); \cite{Iwa16}, 1.27 (SM, Tokyo); \cite{Bar15}, 1.269 (IBM-2); \cite{Vaq13}, 1.25 (GCM, Madrid); \cite{Yao15}, 1.254 (GCM, Sendai); \cite{Men09}, 1.25 (SM, Madrid); \cite{Jia17}, 1.27 (GCM, Chapel Hill); \cite{Ter18}, 1.27  (QRPA, our calculation). 
$^{130}$Te, \cite{Bar15}, 1.269 (IBM-2); \cite{Sim13}, 1.27 (QRPA, T\"{u}bingen); \cite{Mus13}, 1.25  (QRPA, Chapel Hill); \cite{Yao15}, 1.254 (GCM, Sendai); \cite{Vaq13}, 1.25 (GCM, Madrid); \cite{Men09}, 1.25 (SM, Madrid); \cite{Nea15}, 1.254 (SM, Mount Pleasant); \cite{Hyv15}, 1.26 (QRPA, Jyv\"{a}skyl\"{a});   this paper, 1.27 (QRPA, our calculation). 
$^{136}$Xe, \cite{Bar15}, 1.269 (IBM-2); \cite{Sim13}, 1.27 (QRPA, T\"{u}bingen); \cite{Mus13}, 1.25  (QRPA, Chapel Hill); \cite{Yao15}, 1.254 (GCM, Sendai); \cite{Vaq13}, 1.25 (GCM, Madrid); \cite{Men09}, 1.25 (SM, Madrid); \cite{Hor16}, 1.254 (SM, Mount Pleasant); this paper, 1.27 (QRPA, our calculation). }
\end{figure}

Our calculated NMEs and $R^{(0\nu)}_{1/2}$ are summarized in Tables \ref{tab:NMEs_R_0v} and \ref{tab:NMEs_R_0v_1.27}, and 
$R^{(0\nu)}_{1/2}$'s of different groups are compared in Fig.~\ref{fig:reduced_half-life} ($g_A$ = 1.25$-$1.27) together with those of $^{48}$Ca that we calculated previously. 
It is usual to use the bare value of $g_A$ around 1.26 for comparison of different calculations, e.g., \cite{Eng17}. 
Our values, solid red circles in Fig.~\ref{fig:reduced_half-life}, are always low but are not the lowest in the distribution and are close to those of IBM-2. 
The possible overestimation of the $0\nu\beta\beta$ NME by 50 \% mentioned before corresponds to the possible underestimation of $R^{(0\nu)}_{1/2}$ by a factor of 1/2.25. 
The method of Ref.~\cite{Mus13} to calculate the QRPA wave functions is close to our method, however, the corresponding $R^{(0\nu)}_{1/2}$'s are rather different in $^{130}$Te and $^{136}$Xe. It is speculated that the difference comes from the different Skyrme interactions used in those calculations. 

The IBM-2 \cite{Bar15} was used with two values of $g_A$: the unquenched value (the authors use 1.269, see Fig.~\ref{fig:reduced_half-life}),  and their effective values of ``maximal quenching" (the result is not shown in the figure). These effective values are determined by their simple $A$-dependent formula obtained from the $2\nu\beta\beta$ decays; they are 0.528 ($^{130}$Te) and 0.524 ($^{136}$Xe). These values are close to our effective $g_A$ in Table \ref{tab:NMEs_R_0v}. 

The authors of Ref.~\cite{Pir15} picked up triplets of nuclei connected by the $\beta^-$ or $\beta^+$ decay or the electron capture from the mass region of $A=100$$-$$134$ and found that the geometric means of the two calculated NMEs for the ``left-middle" and ``right-middle" combinations of nuclei in those triplets are rather independent of the strength of the pairing interaction (scaled by $g_{pp}$ in their notation), if $g_{pp}$ is not large. On the basis of this interesting discovery, they determined the effective $g_A$ so as to reproduce the experimental mean values, and a simple fitting formula of $g_A$ was obtained as a function of $A$. If this formula can be applied to $A=136$, their effective value is 0.833. The $0\nu\beta\beta$ NME is not shown in that paper.

An early example of an effective $g_A$ value smaller than one is shown in Ref.~\cite{Fae08}. The conclusion of this reference is that $g_A < 1$ is necessary for reproducing the experimental lifetimes to both the single-charge-change and the $\beta\beta$ decays by the pnQRPA. An effective $g_A$ value of 0.39 for $^{128}$Te was also included in the first version of that reference \cite{Fae07a}.  


Recently, the authors of Ref.~\cite{Sim18b} suggested a condition to determine $g_{pp}$ using the $2\nu\beta\beta$ NME with the closure approximation and the spin and isospin symmetries. They obtained a  quenching factor to $g_A$ of 0.712 using the averaged ratio of  the $2\nu\beta\beta$ NME calculated without the closure approximation and the  experimental one with respect to $A$. Their effective $g_A$ value is 1.00, which is one of the values that they have used; for their previous calculations, see Ref.~\cite{Sim18} and references therein. For comprehensive recent review on the effective $g_A$, see Refs.~\cite{Suh17b, Eng17}. 

\begin{table}
\caption{\label{tab:NMEs_R_0v} NMEs and $R^{(0\nu)}_{1/2}$ of the $0\nu\beta\beta$ decay of $^{136}$Xe and $^{130}$Te with the $g_A$ obtained by fitting the experimental $T^{(2\nu)}_{1/2}$. }
\begin{ruledtabular}
\begin{tabular}{cccccc}
Initial & \multirow{2}{*}{ $M^{(0\nu)}$ } & \multirow{2}{*}{ $M^{(0\nu)}_{GT}$ } & \multirow{2}{*}{ $M^{(0\nu)}_{F}$ } &  \multirow{2}{*}{$g_A$} & $R_{1/2}^{(0\nu)}$ \\
\raisebox{3pt}{nucleus}& &  &  &  & ($10^{13}$MeV$^2\,$yr)\\
\hline\\[-10pt]
$^{136}$Xe & $5.040$ & $3.094$ & $-0.467$ & $0.49$ & $1.223$ \\
\multirow{2}{*}{$^{130}$Te} & $6.691$ & \multirow{2}{*}{$3.613$} & \multirow{2}{*}{$-0.709$} & $0.48$ & $0.772$ \\
& $5.400$ & & & $0.63$ & $0.400$ 
\end{tabular}
\end{ruledtabular}
\end{table}
%
\begin{table}
\caption{\label{tab:NMEs_R_0v_1.27} $M^{(0\nu)}$ and $R_{1/2}^{(0\nu)}$ for $g_A=1.27$. The closure approximation is used for $M^{(0\nu)}$, thus we have only one NME and $R_{1/2}^{(0\nu)}$ for $^{130}$Te. }
\begin{ruledtabular}
\begin{tabular}{ccc}
Initial & \multirow{2}{*}{ $M^{(0\nu)}$ } & $R_{1/2}^{(0\nu)}$ \\
\raisebox{3pt}{nucleus}& & ($10^{13}$MeV$^2\,$yr)\\
\hline\\[-10pt]
$^{136}$Xe & $3.109$ & $0.075$ \\
$^{130}$Te & $3.725$ & $0.054$
\end{tabular}
\end{ruledtabular}
\end{table}
%

\begin{figure}[]
\includegraphics[width=1.0\columnwidth]{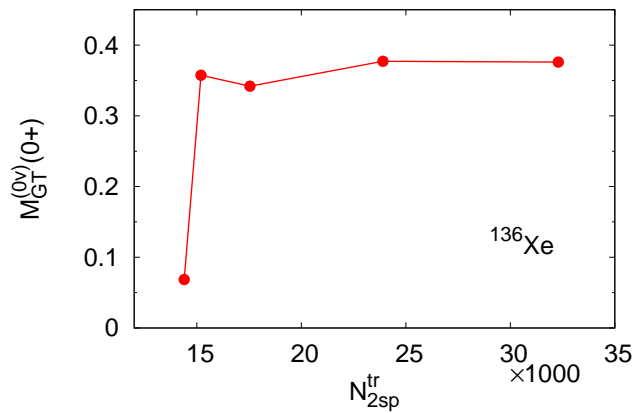}
\caption{ \label{fig:numtr_nme_xe136-ba136}  GT component ($K^\pi = 0^+$) of $M^{(0\nu)}$ of $^{136}$Xe $\rightarrow$ $^{136}$Ba as a function of $N^\mathrm{tr}_\mathrm{2sp}$, which is the summation of the dimensions of the two-single-particle (the canonical basis) states used for the transition from $|a^0_F\rangle$ to $|F\rangle$ and those from $|I\rangle$ to $|a^0_I\rangle$; see Eqs.~(\ref{eq:M0vGT}) and (\ref{eq:M0vF}). }
\end{figure}

\begin{figure}[]
\includegraphics[width=1.0\columnwidth]{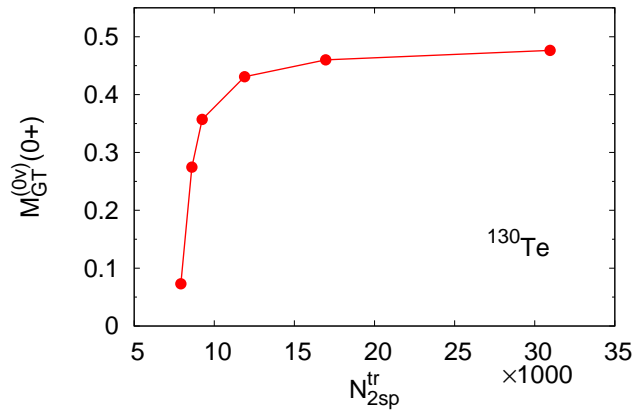}
\caption{ \label{fig:numtr_nme_te130-xe130}  The same as Fig.~\ref{fig:numtr_nme_xe136-ba136} but for $^{130}$Te $\rightarrow$ $^{130}$Xe. }
\end{figure}

Figure \ref{fig:numtr_nme_xe136-ba136} shows the convergence of the NME of $^{136}$Xe $\rightarrow$ $^{136}$Ba with respect to the single-particle space. 
We use very large two-quasiparticle spaces for the QRPA calculations, and the test of the convergence was made by varying the two-single-particle spaces used for calculating $V^{GT(0\nu)}_{pp^\prime,nn^\prime}(\bar{E}_a)$ and 
$V^{F(0\nu)}_{pp^\prime,nn^\prime}(\bar{E}_a)$. The well-converged results were used for the calculation of the NMEs. 
The behavior of the convergence of $^{136}$Xe is less smooth than that of $^{130}$Te (Fig.~\ref{fig:numtr_nme_te130-xe130})  because the ground states of $^{136}$Xe and $^{136}$Ba are spherical, and the neutrons of $^{136}$Xe are unpaired; see Table \ref{tab:HFB_solution}. 

\subsection{\label{sec:charge-change} Charge-change reaction}
\begin{figure}[]
\includegraphics[width=1.0\columnwidth]{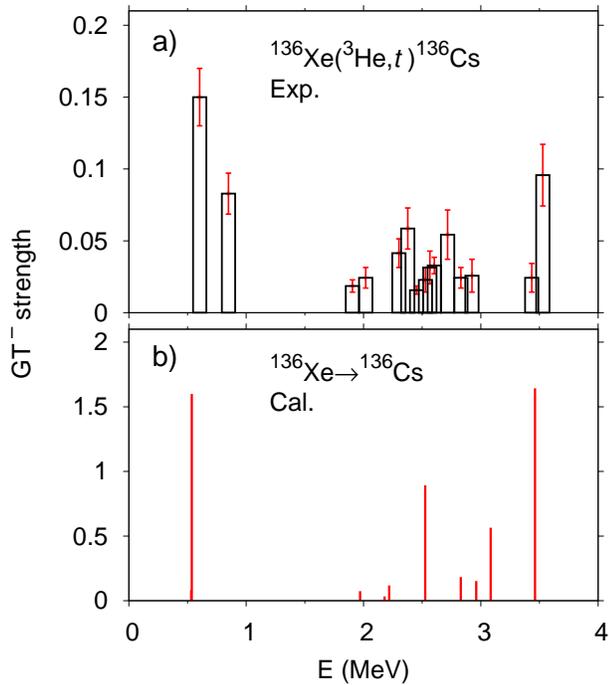}
\caption{ \label{fig:gtstr_136Xe} GT$^-$ transition strength of $^{136}$Xe obtained by experiment (a) and our calculation (b). 
$E$ is the excitation energy of $^{136}$Cs. The data were obtained from Ref.~\cite{Pup11}. }
\end{figure}

There are experimental data of $^{136}$Xe($^3$He,\textit{t})$^{136}$Cs \cite{Pup11,Fre17}, that can be used for test of the charge-change transition density 
$\langle a^K_I | c_{p^\prime}^\dagger c_{n^\prime} | I \rangle$ [see Eqs.~(\ref{eq:M0vGT}) and (\ref{eq:M0vF})]   
independently of $g_A$. 
The experimental and calculated GT$^-$ strengths are shown in Fig.~\ref{fig:gtstr_136Xe} for a low-energy region in which the GT$^-$-strength  data are obtained. 
The calculated one is the summation with respect to $K$ of the squared transition matrix element $|M_\mathrm{GT}^K|^2$ of the GT$^-$ operator $\sigma\tau^-$, and the data are obtained from the cross section.  
The two results are similar in terms of the energy dependence, however, the calculated values are larger than the data by one order of magnitude.  
The GT sum-rule value of our calculation is 
$$
85.145 - 1.138 = 84.007, 
$$
where the first term is the GT$^-$ component, and the second term is the GT$^+$ component ($^{136}$Xe $\rightarrow$ $^{136}$I). The sum rule is satisfied well, i.e., our value is very close to $3(N-Z)$. 

There was a similar problem in $^{48}$Ca $\rightarrow$ $^{48}$Sc and $^{48}$Ti $\rightarrow$ $^{48}$Sc \cite{Yak09}. This problem was solved \cite{Ter18} by using the transition operator consisting of the GT and isovector spin-monopole operators phenomenologically. 
This idea can be tested, if the GT-strength data including the giant-resonance region are obtained. Actually the yield data are obtained up to 30 MeV. For the shell-model approach to these data, see Ref.~\cite{Nea15}.

\begin{figure}[]
\includegraphics[width=1.0\columnwidth]{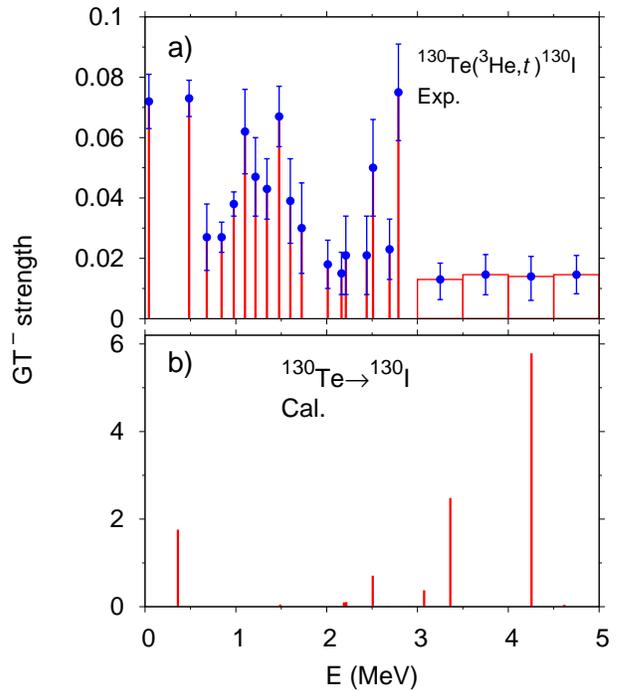}
\caption{ \label{fig:gtstr_130Te} GT$^-$ transition strength of $^{130}$Te obtained by experiment (a) and our calculation (b). 
$E$ is the excitation energy of $^{130}$I. The data were obtained from Ref.~\cite{Pup12}. 
Above 3 MeV in panel a the transition strengths are integrated over the 500-keV intervals and distributed into 35-keV bins \cite{Pup12}.}
\end{figure}

We show in Fig.~\ref{fig:gtstr_130Te} the GT$^-$ strength obtained from the reaction of $^{130}$Te($^3$He,\textit{t})$^{130}$I \cite{Pup12} and that obtained by our calculation. Contrary to the GT$^-$ strength of $^{136}$Xe$\rightarrow$$^{136}$Cs, we have much fewer $1^+$ states than the data have. The deformation effect discussed in Sec.~\ref{sec:spectrum} is also reflected here. (The data show more $1^+$ states than the compilation of Ref.~\cite{nndc}.)  
The total strength up to 5 MeV is 1.5 \cite{Pup12} and 11.4 in our calculation, thus, the same argument as  that for $^{136}$Xe$\rightarrow$$^{136}$Cs is applied here; 
the experimental strength distribution in a very large energy region is necessary for discussing the possible necessity of the isovector spin-monopole operator. 
Our GT sum-rule value is 77.951, and the exact value is 78.
%

\subsection{\label{sec:beta}$\bm{\beta}$ decay}
\begin{figure*}[]
\includegraphics[width=2.0\columnwidth]{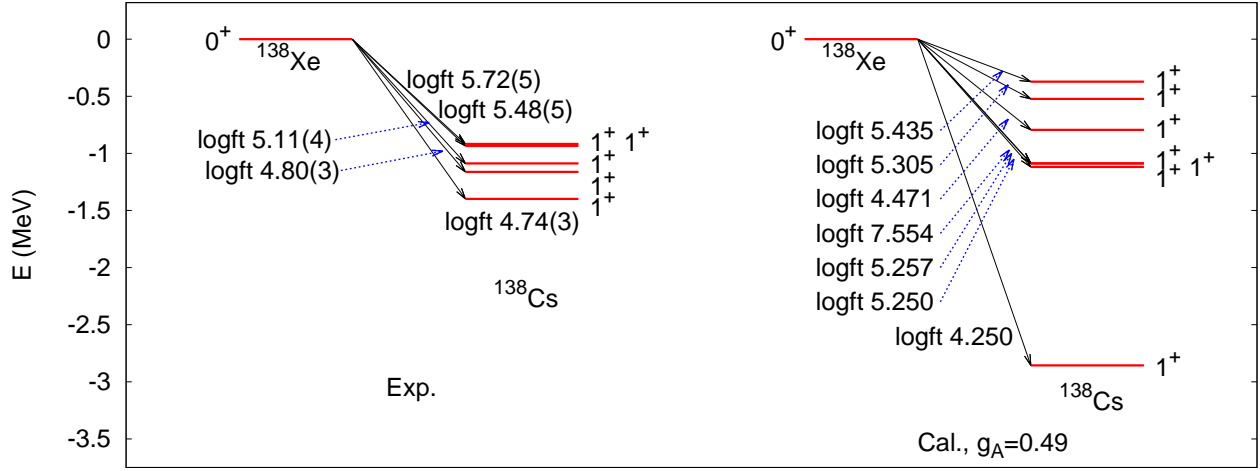}
\caption{ \label{fig:bdecay_138Xe} Spectrum of final $1^+$ states of $\beta$ decay of $^{138}$Xe and logft. Left (right) is the experimental data \cite{nndc} (our calculation). }
\end{figure*}

\begin{figure*}[]
\begin{minipage}{1.0\columnwidth}
\includegraphics[width=1.0\columnwidth]{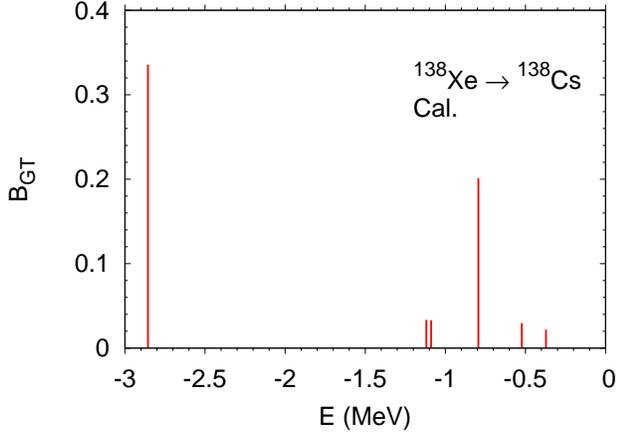}
\caption{ \label{fig:bgtcal_138Xe} Calculated $B_\mathrm{GT}$ of $\beta$ decay of $^{138}$Xe as a function of the energy of $^{138}$Cs relative to the ground state of $^{138}$Xe. }
\end{minipage}
\begin{minipage}{1.0\columnwidth}
\includegraphics[width=1.0\columnwidth]{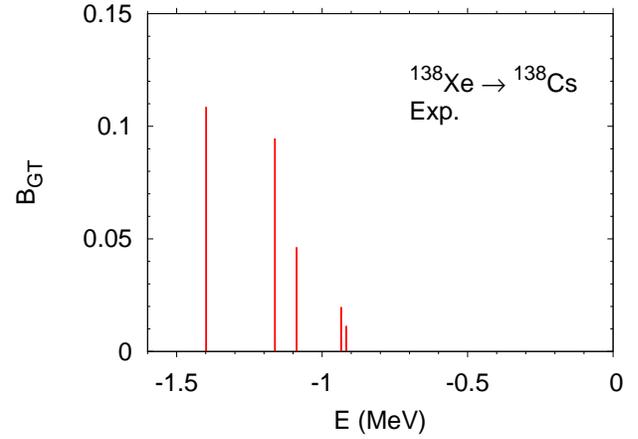}
\caption{ \label{fig:bgt_138Xe} The same as Fig.~\ref{fig:bgtcal_138Xe} but for experimental data \cite{nndc}. }
\end{minipage}
\end{figure*}

We calculated the probability of the $\beta$ decay of $^{138}$Xe for testing our $g_A$. The calculated spectrum and logft are compared to those of the experimental data \cite{nndc} for $1^+$ states in Fig.~\ref{fig:bdecay_138Xe}. The level distribution of the data is slightly denser and lower than the corresponding calculated levels. We obtained a level around $-$2.9 MeV, however, there is no corresponding experimental level. Actually there is a level of $0^-$ or $1^-$ around that energy in the data. We plot 
$$ 
B_\mathrm{GT} = g_A^2 \sum_{K=0,\pm 1} |M_\mathrm{GT}^K|^2 ,
$$ 
in Figs.~\ref{fig:bgtcal_138Xe} (our calculation) and \ref{fig:bgt_138Xe} (experiment). The summation of the calculated $B_\mathrm{GT}$ between $-1.2$ MeV and $-0.3$ MeV is 0.319, and the corresponding experimental value is 0.281; the former is 14 \% larger than the latter. Thus, our $g_A$ fitting the $T^{(2\nu)}_{1/2}$ of $^{136}$Xe is consistent with the $\beta$ decay of a nearby nucleus. 

\begin{figure*}[]
\includegraphics[width=2.0\columnwidth]{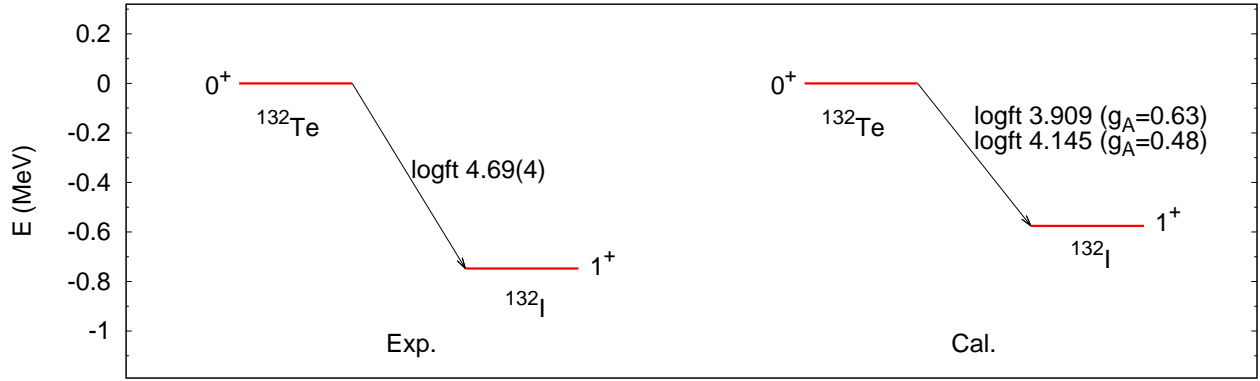}
\caption{ \label{fig:bdecay_132Te} The same as Fig.~\ref{fig:bdecay_138Xe} but for $^{132}$Te.}
\end{figure*}

The GT$^-$ spectrum of $^{132}$Te $\rightarrow$ $^{132}$I is shown in Fig.~\ref{fig:bdecay_132Te} with the logft values. Only one $1^+$ final state is observed experimentally \cite{nndc}. The logft by our calculation is smaller than the data by $\simeq$ 0.5$-$0.8, and the corresponding $B_\mathrm{GT}$ ratio is 3.2$-$6.3. The quality of this calculation is not as good as that for $^{138}$Xe. 


\subsection{ \label{sec:gt3} Higher-order effect in 2$\bm{\nu\beta\beta}$ decay }
Very recently, a higher-order component of the $2\nu\beta\beta$ NME \cite{Sim18}
\begin{eqnarray}
\frac{ M^{(2\nu)}_{GT\textrm{-}3} }{ \mu_0 } &=& \sum_{K=0,\pm 1} \sum_{a^K_I, a^K_F}
 \frac{4}{ \mu_a^3 } \langle F | \tau^- (-)^K \sigma_{-K} | a^K_F\rangle \nonumber \\
&&\times \langle a^K_F | a^K_I \rangle \langle a^K_I | \tau^- \sigma_K | I \rangle , \label{eq:M2vGT3}
\end{eqnarray}
was suggested, and the experimental value of the ratio 
\begin{eqnarray}
\xi_{31}^{(2\nu)} = \frac{ M^{(2\nu)}_{GT\textrm{-}3} }{ M^{(2\nu)}_{GT} },
\end{eqnarray}
was extracted for $^{136}$Xe by fitting the $2\nu\beta\beta$ spectrum \cite{Gan19}. 
In Eq.~(\ref{eq:M2vGT3}), $\mu_a$ refers to Eq.~(\ref{eq:mean_M(I)}) or (\ref{eq:mean_M(F)}); there is no difference for $^{136}$Xe. 
Our value is $\xi^{(2\nu)}_{31} = 0.040$, which is within the experimentally reported range of 
$\xi^{(2\nu)}_{31} = -0.26^{+0.31}_{-0.25}$ determined by $1\sigma$ for exclusion.  
This comparison also enhances the validity of our calculation. 
Because this discussion is very new, $M^{(2\nu)}_{GT\textrm{-}3}/\mu_0$ is not included in the $2\nu\beta\beta$ NME in Sec.~\ref{sec:2v}.  

\section{ \label{sec:summary} Summary }
In this paper, we have calculated the NMEs of the $\beta\beta$ decays of $^{136}$Xe and $^{130}$Te, from which $R^{(0\nu)}_{1/2}$'s were derived. Several tests have been conducted to investigate the reliability of our calculation of $R^{(0\nu)}_{1/2}$:  
\begin{enumerate}
\item 
the self-check of the dual intermediate states using the $2\nu\beta\beta$ decay.

\item
The convergence of the NME with respect to the dimension of the single-particle space.

\item
The comparison of the two spectra of the intermediate nucleus obtained by the pnQRPA and comparison of them with the experimental data. 

\item
The GT sum rule.

\item
The comparison of the GT$^-$ strength with the data of the charge-change reaction.

\item
The comparison of the $\beta$ decay spectrum and $B_\mathrm{GT}$ with the experimental data. 

\item
Test using a new quantity expressing a higher-order effect in the $2\nu\beta\beta$ NME. 
\end{enumerate}
No problem was found for $^{136}$Xe concerning our methodology. On the other hand, it turned out that our approach was not successful for $^{130}$Te. 

A key point of our procedure is the strength of the isoscalar pairing interaction. Our value is not very large, so that the pnQRPA solutions are not close to the unstable region. Actually, this value makes the effective $g_A$ smaller. Thus, we investigated if this $g_A$ was consistent with the $\beta$ decay, and the consistency was obtained. The effective $g_A$ depends on the approximation. To our knowledge, the spectrum of the intermediate nucleus by the pnQRPA has been shown for the first time. The comparison of the calculated GT$^-$ strength with the experimental data ($^{136}$Xe) is encouraging in terms of the energy dependence, however, a problem remains about how the experimental absolute value can be reproduced; a possibility is to include the isovector spin-monopole operator in the transition operator. The GT$^-$-strength data in a broader energy region are necessary for the test of the calculation. That transition operator is not obvious a priori in the charge-change reaction. 


\begin{acknowledgments} 
The numerical calculations of this paper were performed by 
the K computer at RIKEN Center for Computational Science through the program of High Performance Computing Infrastructure in 2017B (hp170288) and 2018B (hp180232). Computer Oakforest-PACS operated by Joint Center for Advanced High Performance Computing was also used through 
Multidisciplinary Cooperative Research Program of Center for Computational Sciences, University of Tsukuba in 2018 (xg18i006). Oakforest-PACS has also been used through High Performance Computing Infrastructure (hp190001, 2019).
This study is supported by 
European Regional Development Fund, Project ``Engineering applications of microworld physics" (No.~CZ.02.1.01/0.0/0.0/16\_019/0000766). 
\end{acknowledgments}


\appendix 

\section{\label{sec:appendix1} Identification of nucleus in QRPA solutions}
The QRPA solutions include multiple nuclei, if the HFB ground state is paired. In the low-energy region, the identification of the solutions corresponding to a specific nucleus is possible only approximately. We introduce the auxiliary transition operator $O^-_\mathrm{id}(J^\pi)$ for this identification: 
\begin{eqnarray}
O^-_\mathrm{id}(0^+) &=& \frac{ r^8 }{ \langle r^8 \rangle } \tau^-,
\end{eqnarray}
\vspace*{-12pt}
\begin{eqnarray}
O^-_\mathrm{id}(0^-) &=& \frac{ r^8 }{ \langle r^8 \rangle } \tau^- [ Y_1 \sigma ]_{00} , 
\end{eqnarray}
and those for other $J^\pi$ in the analogous way. We calculate the transition strengths of these operators for the QRPA solutions based on $^{136}$Xe and regard those having relatively large strengths as the states of $^{136}$Cs. The $J^\pi$ of the state is identified with that of the transition operator of the large strength.
The expectation value of $r^8$ is calculated with respect to the HFB ground state. 
This $r^8$ is used for covering a large energy region, as seen by expressing it using the creation and annihilation operators of the harmonic oscillator. The exponent of 8 may be reasonable because $K$ values up to 8 are treated in the NME calculation of the $\beta\beta$ decay. 

For identifying $^{136}$Cs obtained from $^{136}$Ba, the same operators but for $\tau^+=\tau^{-\dagger}$ instead of $\tau^-$ are used. 
In neutron-rich nuclei, the $p\rightarrow n$ transition is more rare compared to the $n \rightarrow p$ transition. Therefore,  all QRPA solutions that have the finite transition strength of $p \rightarrow n$ were picked up. 

\section{\label{sec:appendix2} Assignment of $\bm{J}$ to deformed QRPA solutions}
We modify slightly the auxiliary transition operator in Appendix \ref{sec:appendix1} for assigning major $J$ to the deformed states obtained from $^{130}$Xe (Sec.~\ref{sec:spectrum}). 
If $J^\pi$ is of the unnatural parity, 
\begin{eqnarray}
O^+_\mathrm{Jid}(J^\pi) &=& \frac{1}{{\cal S}(J^\pi)}r^2 \tau^+ [ Y_{J-1} \sigma ]_{JK}, \ \ (J\geq 1), \\
O^+_\mathrm{Jid}(0^-) &=& \frac{1}{{\cal S}(J^\pi)}r^2 \tau^+ [ Y_1 \sigma ]_{00}, 
\end{eqnarray}
is used. ${\cal S}(J^\pi)$ is the factor normalizing the sum of the transition strengths for different $J^\pi$'s. The transition strengths of this operator are calculated for all QRPA solutions ($K\le 8$) with the specified parity in a low-energy region, and the most and next-most major $J$'s are noted in panel b of  Fig.~\ref{fig:spectrum_130I_cal}. This method was confirmed to give the correct $J$ for the spherical states of $^{130}$Te; the correct $J$ is seen by the degeneracy. This check is not trivial because our calculation uses the single-particle wave functions in the cylindrical box with the  vanishing boundary condition. 
For the natural-parity states,  
\begin{eqnarray}
O^+_\mathrm{Jid}(J^\pi) &=& \frac{1}{{\cal S}(J^\pi)} r^2 \tau^+ Y_{JK},
\end{eqnarray}
is used. 

\bibliography{xe136}
\end{document}